# Growth of macroscopic-area single crystal polyacene thin films on arbitrary substrates

Randall L. Headrick[1], Hua Zhou[1], Binran Wang[1], Yiping Wang[1], Greggory P. Carpenter[1], Alex C. Mayer[2], Matthew Lloyd[2], George G. Malliaras[2], Alexander Kazimirov[3], and John E. Anthony[4]

[1) Department of Physics, University of Vermont, Burlington, Vermont 05405, USA

2) Materials Science and Engineering, Cornell University, Ithaca, New York 14853, USA

3) Cornell High Energy Synchrotron Source, Ithaca New York, 14853, USA

4) Department of Chemistry, University of Kentucky, Lexington, Kentucky 40506, USA

Organic electronic materials have potential applications in a number of low-cost, large area electronic devices such as flat panel displays and inexpensive solar panels. Small molecules in the series Anthracene, Tetracene, Pentacene, are model molecules for organic semiconductor thin films to be used as the active layers in such devices.[1-5] This has motivated a number of studies of polyacene thin film growth and structure.[6-12] Although the majority of these studies rely on vapor-deposited films, solvent-based deposition of films with improved properties onto non-crystalline substrates is desired for industrial production of devices. Improved ordering in thin films will have a large impact on their electronic properties, since grain boundaries and other defects are detrimental to carrier mobilities and lifetimes. Thus, a long-standing challenge in this field is to prepare large-area single crystal films on arbitrary substrates. Here we demonstrate a solvent-based method to deposit thin films of organic semiconductors, which is very general. Anthracene thin films with single-crystal domain sizes exceeding $1 \times 1$ cm$^2$ can be prepared on various substrates by the technique. Films of 6,13-bis(triisopropylsilylethynyl)pentacene are also demonstrated to have grain sizes larger than $2 \times 2$ mm$^2$. In contrast, films produced by conventional means such as vapor deposition or spin coating are polycrystalline with micron-scale grain sizes. The general propensity of these small molecules towards crystalline order in an optimized solvent deposition process shows that there is great **potential for thin films with improved properties. Films prepared by these methods will also be useful in exploring the limits of performance in organic thin film devices.**

Vapor deposition is the standard technique to form most organic semiconductor thin films. However, we use an alternative approach borrowed from recent research on the crystallization of synthetic spheres from a colloidal suspension.[13,14] It has been shown that strong capillary forces at the meniscus between a substrate and a colloidal suspension of spherical particles can induce crystallization into a nearly two-dimensional array of controllable thickness. If this meniscus is slowly swept across a vertically placed substrate by solvent evaporation, thin two-dimensional opals can be deposited.[14] We have discovered that this process is also an effective method for the growth of large-area crystalline films from solution (A schematic shown in the Supplementary Information).

Anthracene has a monoclinic structure with lattice constants $a = 8.561$ Å, $b = 6.036$ Å, $c = 11.163$ Å and $\beta = 124°$ $42$'.[15] The structure is composed of layers of molecules stacked along the the c-direction with the "herringbone" packing within each layer (Supplementary Figure 2). The (001) surface has the lowest free energy, and as a result, when a thin film of anthracene is formed the **a** and **b** lattice vectors are typically in the plane of the film. Individual molecules in the film stand nearly upright with respect to the surface, but "lean over" by an angle $\chi = (\beta - 90) = 34.6°$ from the surface normal. The crystal structure of 6,13-

bis(triisopropylsilylethynyl)pentacene is reported to be triclinic with $a = 7.565$ Å, $b = 7.750$ Å, $c = 16.835$ Å, $\alpha = 89.15°$, $\beta = 78.42°$, and $\gamma = 83.63°$.[16]  The molecules do not pack into the herringbone arrangement of the parent compound, but rather crystallize into a 2D "bricklayer" arrangement, which maximizes $\pi$-orbital overlap between the aromatic cores on adjacent molecules. Transistors fabricated from thin films of this material exhibit hole mobilities as high as 0.4 cm$^2$/V-s.[17,18]

We have obtained excellent results for anthracene deposition on a variety of substrates, including glass, oxidized silicon, and polymers.  We find that a variety of film morphologies are formed including continuous films and separated wire-like structures with individual widths as small as a few microns.  Single-crystal domains approach the length of the sample in one direction (we used samples up to 75 mm in our experiments).  A practical advantage of this method is the ability to coat relatively large areas easily without resorting to a vacuum environment.

In one trial run, a glass microscope slide was placed vertically in a vial containing a solution of anthracene in ethyl acetate.  As the solution slowly evaporated, the meniscus of the solution was swept across the surface towards the lower edge of the slide, depositing a thin film of anthracene.  The anthracene film produced by this method is colorless and perfectly transparent.  Figure 1 shows a photograph of the film viewed through crossed polarizers with illumination from behind the sample.  The anthracene film induces a large rotation of the polarization producing the contrast.  As a result, a clear pattern of domains becomes visible that were not readily apparent under normal room lighting.

Control over the thickness and morphology of films can be achieved by varying the concentration of the solution.  However, we have chosen a more novel method to vary the structure of our films.  Draining or pumping away the solvent at a controlled rate achieves direct control over the *in-plane* growth rate.  Growth rates larger than 1 cm/hr typically produce films with submicron thickness.  However, films become discontinuous for in-plane growth rates larger than about 2 cm/hr.  For example, figure 2 shows a fluorescence micrograph of an anthracene film deposited at ~5 cm/hr.  The bright areas are anthracene and the dark areas are the silicon underneath. The image shows that an unexpected two-dimensional rowlike structure is formed.  Additional imaging by polarized light microscopy show that the crystallographic orientation is the same over the entire area shown ($1.25 \times 1$ mm$^2$).

The apparent mechanism that forces the selection of the highly oriented domains shown in Figures 1 and 2 is rather interesting.  Small nuclei form early in the process and become elongated as the liquid level is lowered down the surface of the substrate.  A preferred crystallographic direction is selected, since crystallites grow faster in certain low-index growth directions.  Slow-growing nuclei with unfavorable orientations are left behind as the process proceeds, and the fast-growing domains increase rapidly in width, eventually squeezing out less favored orientations.  The end result is the domain structure shown in Figures 1 and 2, where single-crystal domains stretch almost the entire length of the surface in the growth direction, and may exceed one centimeter in the transverse direction. Adjacent domains have similar orientations, lying within a range of approximately ±10°. The mechanism is similar to the one proposed to explain recent observations of grain growth in ion beam assisted deposition of inorganic thin films, where faster growing grains with preferred crystallographic orientations overtake and shadow slower-growing domain orientations.[19]

Figure 3a shows the results of an x-ray diffraction θ-2θ scan of the sample shown in figure 2. Six orders of (00L) reflections are clearly observed, indicating that the film is of good quality with the c* reciprocal lattice axis oriented normal to the surface. A layer spacing of d = 9.18 Å was derived from this scan, which is consistent with the crystal structure of bulk anthracene.

Grazing incidence x-ray diffraction scans were performed on additional anthracene films, in order to establish the in-plane orientation of the layers. A sample prepared in the same manner as Fig. 2 exhibited a single-crystal rowlike surface morphology and was found to be predominantly (>95%) single crystal by polarization microscopy. Figure 3b shows an azimuthal scan on the (200) reflection of anthracene for this sample. If the film were polycrystalline, composed of domains with a perfectly random distribution of azimuthal orientations, a continuous low intensity would be observed. Since we observe one dominant reflection at 0°, the film is predominantly single-crystal. However, the right inset shows evidence for two grains with ~1° misorientation relative to each other.

In order to assist the reader in interpreting the orientations of the anthracene crystal planes in the aligned samples, we have included a stereographic projection as an inset to Fig. 3b. The polar angle between the (001) reflection and the (200) reflection is $\beta^* = 180 - \beta$, where $\beta$ is the real-space angle between the **a** and **c** lattice vectors of the thin-film crystal structure, and $\beta^*$ is the corresponding angle for the reciprocal-space unit cell . Similarly, the polar angle between (001) and (020) is 90º. The (200) is observed at $\beta^* = 55.4°$, consistent with the structure of anthracene. The path of the scan corresponding to the (200) data shown in Fig. 3b is indicated in the inset by the red dashed line.

The azimuthal orientation of the long rowlike features of the Anthracene film can also be obtained from Figure 3b. The arrow in the inset of Fig 2 indicates the growth direction. We have oriented the sample so that the scattering vector is perpendicular to the film growth direction at the azimuthal angle ϕ=0; this was accomplished by using a laser pointer to produce a visible light diffraction pattern from the periodic rowlike structure, and then visually orienting the sample on the x-ray diffractometer to an accuracy of about 1°. Therefore, since the (200) reflection appears at ϕ=0, we conclude that the (200) reciprocal lattice direction is perpendicular to the anthracene growth direction. In other words the domains shown in Figure 3a are oriented with their long axes parallel to the [010] crystallographic direction.

The data of Fig. 3b confirm that the individual rowlike features in Fig. 2 have identical crystallographic orientations. Apparently, a combination of the preferred orientation imposed by the deposition process, and the links between adjacent branches in the structure is enough to select a single crystallographic orientation for the whole area. We do not fully understand the origin of this rowlike structure, but it is likely related to well-known growth instabilities such as the Mullins-Sekerka instability,[20] which may be induced by concentration gradients as molecules diffuse in the region of the meniscus and are depleted by incorporated into the film.

Pentacene has a rather low solubility in common solvents. Therefore we demonstrate results for a silylethynyl-substituted derivative of pentacene. Growth morphologies were observed that were analogous to those observed for anthracene, ranging continuous films for slow deposition (Figure 4) to row-like discontinuous films for higher deposition rates (Supplementary Figure 3). Figure 4 shows three bright-field optical micrographs of the same area of a thin film deposited onto a glass slide. Image (a) shows a domain with dimensions of about 2.5 mm × 2.5 mm and its boundary with an adjacent domain. Image (b) shows the same region with crossed polarizers. Note that the adjacent domain turns dark. In image (c), the sample has been rotated by ~50° and the large grain turns dark. The polarization contrast shows that domains are single the crystal. The close correspondence of the results between anthracene and a pentacene derivative illustrates the generality of the solvent/meniscus method as applied to the deposition of polyacene semiconductor thin films.

**Methods**

Anthracene films were deposited from a 2.0 mg/mL solution of anthracene in ethyl acetate. Films of silylethynylated-pentacene were deposited from a 0.25 mg/mL solution of 6,13-bis(triisopropylsilylethynyl)pentacene in ethyl acetate. Two variations of the film deposition method were used: (i) A 75 mm by 25 mm sample of glass or silicon was placed upright in a staining jar (Krackeler Scientific Inc.) containing the solution.  The sample in solution was left undisturbed for several hours, allowing the solvent to evaporate.   (ii) A 600 mL beaker was used as a container, and an oxidized silicon sample was suspended upright using a fixture to hold the sample from the top.  A peristaltic pump (Rainin Instrument Co. Inc.) was used to gradually reduce the solution level at a controlled rate.  All samples were subsequently examined with an optical microscope (Zeiss Axioskop), using polarized-light microscopy and fluorescence microscopy with illumination from a mercury lamp. Selected samples were examined with x-ray diffraction.

X-ray diffraction experiments were performed at the A2 beamline of the Cornell High Energy Synchrotron Source using 10.0 keV x-rays ($\lambda = 1.239$ Å) with a flux of $\sim 10^{12}$ photons/sec, incident to the sample, which was mounted on a four-circle diffractometer. A scintillation counter was used for measuring the scattered x-ray intensity.

**Supplementary Information** has been submitted with this manuscript.

## Acknowledgements

This research was supported through the CAREER program of the National Science Foundation, and through the Cornell Center for Materials Research, a Materials Research Science and Engineering Center of the National Science Foundation. The National Science Foundation also supports the Cornell High Energy Synchrotron Source.

## Competing Interests

The authors declare that they have no competing financial interests.

## Correspondence

Correspondence and requests for materials should be addressed to R.L.H. (rheadrick@uvm.edu).

## Figure Captions

**Figure 1. Unmagnified photograph of an anthracene thin film deposited on a 75 mm by 25 mm glass slide viewed through crossed polarizers.** The light source is behind the sample, followed by the first polarizer (the black disk visible in the image), and the sample itself. The domain structure of the film is visible, with a grain size exceeding one centimeter in some areas of the film. The domains are much longer along the growth direction, which is along the long direction of the slide. Rotation of the sample by 45° reverses the contrast.

**Figure 2. Fluorescence micrograph of single crystal anthracene deposited on an oxidized silicon surface.** During the growth of this film, the meniscus of the solvent was swept towards the bottom of the image. Two-dimensional rowlike structures are observed, which are nearly aligned with the nominal growth direction. Faint horizontal features are observed in the image, which mark the position of the solvent's meniscus at several points during the growth process. They are probably the result of vibration or air currents, which tend to disturb the surface of the solvent. The image area is $1 \times 1.25$ mm$^2$, and the inset shows an enlarged view of the same image. The arrow in the inset indicates the growth direction of the rows, which is found to be along the [010] crystallographic direction.

**Figure 3. X-ray diffraction analysis of an anthracene thin film deposited on an oxidized silicon wafer.** In (a) six orders of (00L) reflections are observed. In (b), x-ray diffraction data are shown for the anthracene (200) reflection. The dominant anthracene crystal orientation results in reflection at $\phi = 0°$. The right inset shows an expanded view of the anthracene data near $\phi = 0°$, revealing two peaks separated by 1°. The left inset shows a stereographic projection, depicting the azimuthal and polar angles of each reflection, and the path of the azimuthal scan (red line).

**Figure 4. Bright-field micrographs of a thin film of silylethynyl-substituted pentacene deposited on a glass slide.** These visible-light images show an area of $1.75$ mm $\times 1.25$ mm of a 6,13-bis(triisopropylsilylethynyl)pentacene film. Image (a) shows a domain with dimensions of about $2.5$ mm $\times 2.5$ mm and its boundary with an adjacent domain. Image (b) shows the same region with crossed polarizers. In (c), the sample has been rotated by ~50° with respect to image (b).

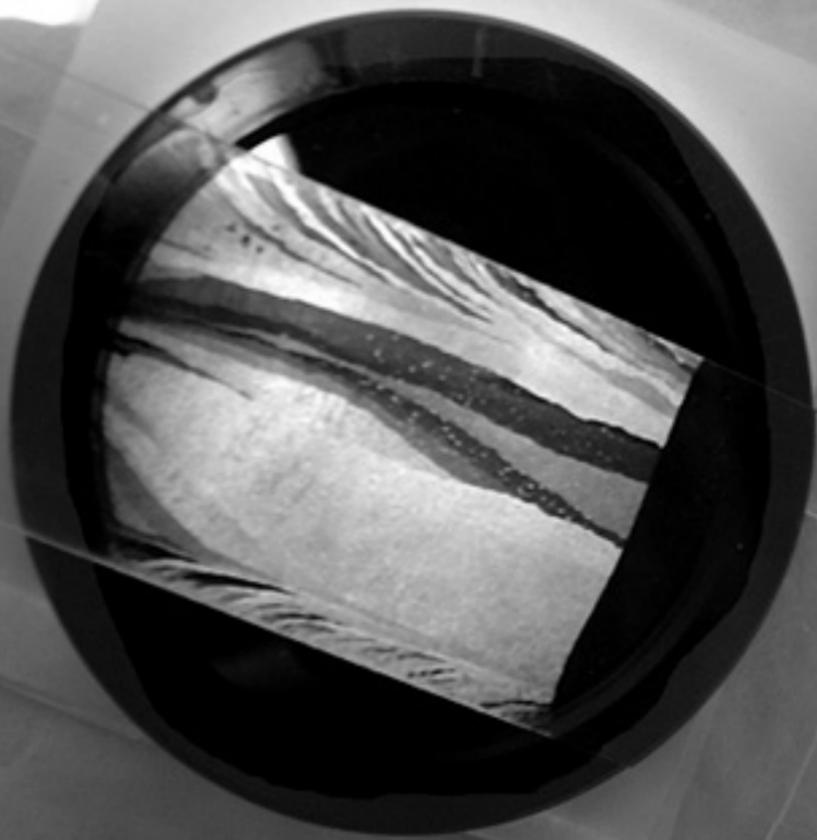

headrick_fig2

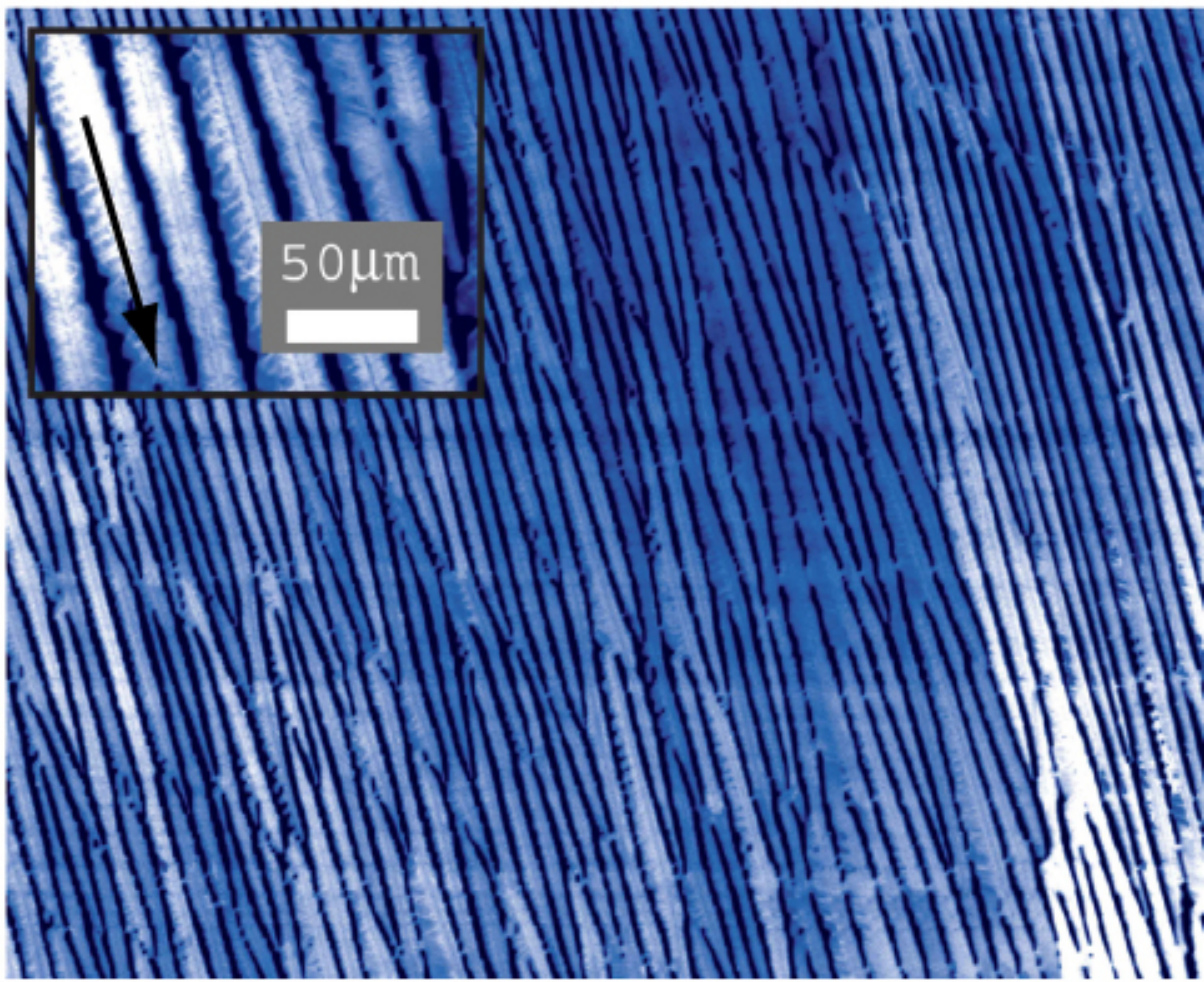

50µm

(a)

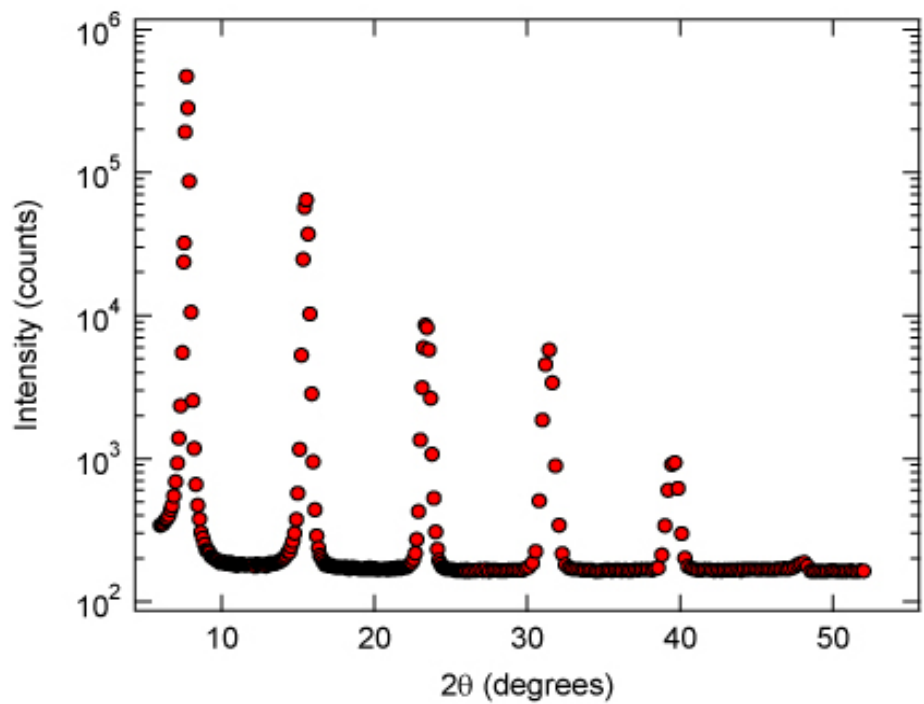

(b)

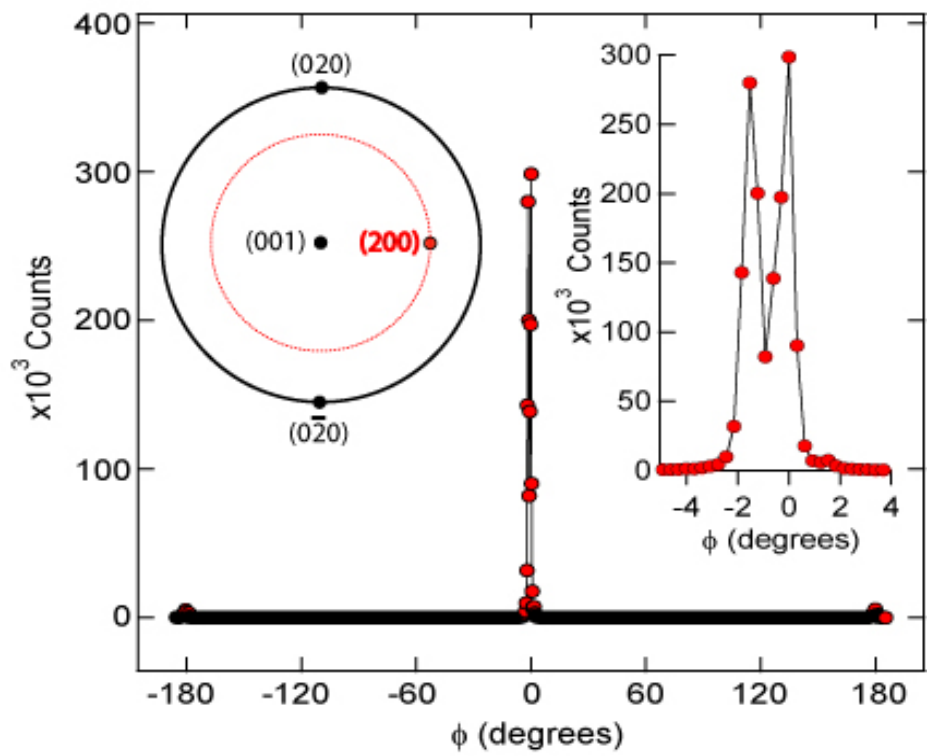

headrick_fig4

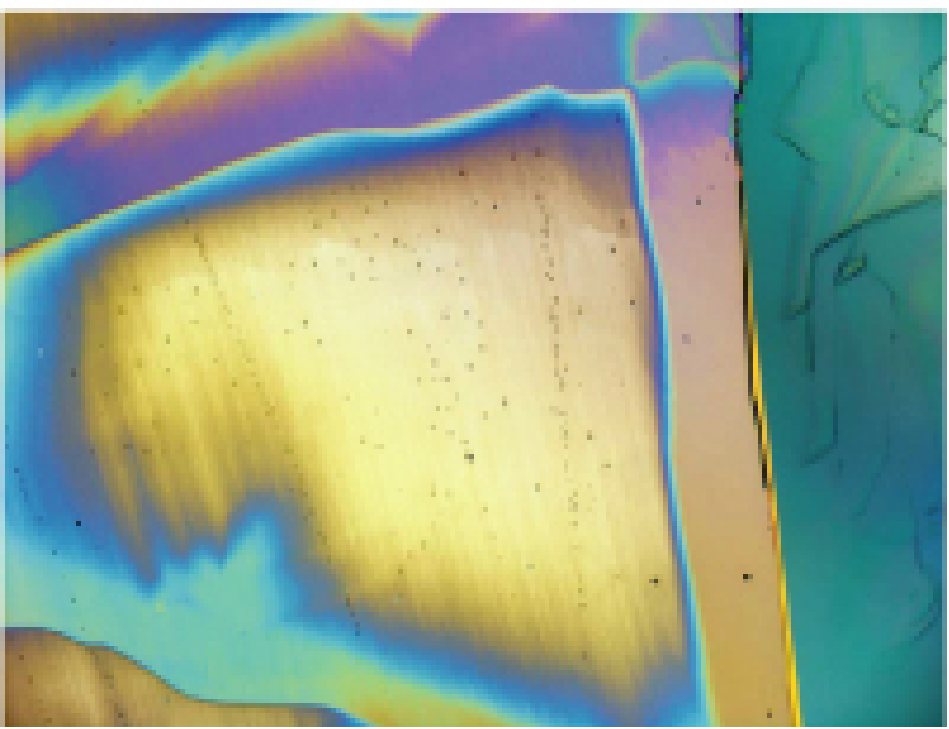 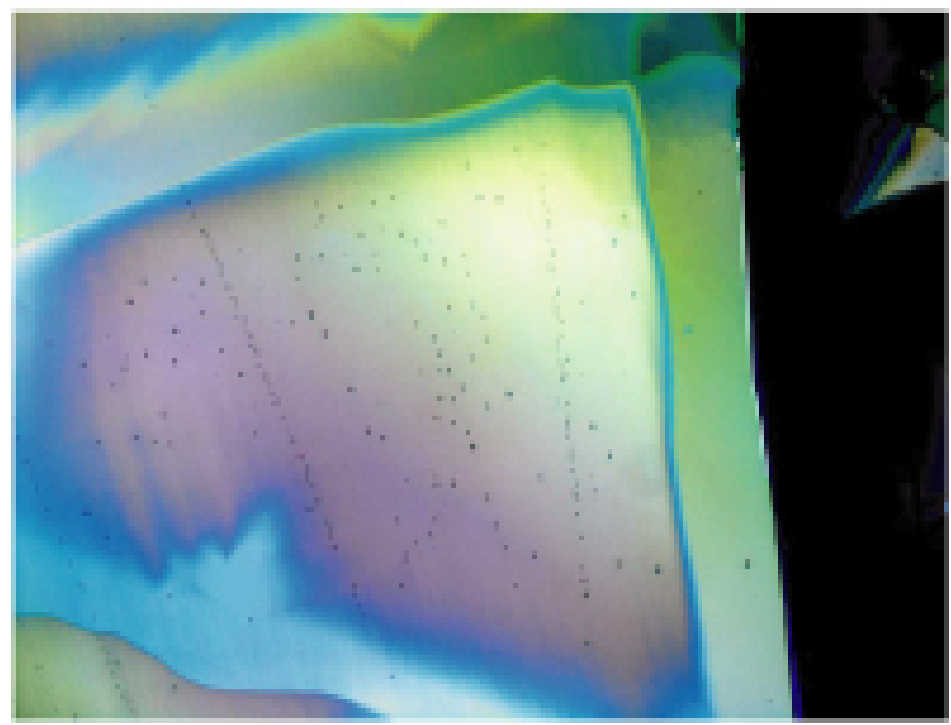 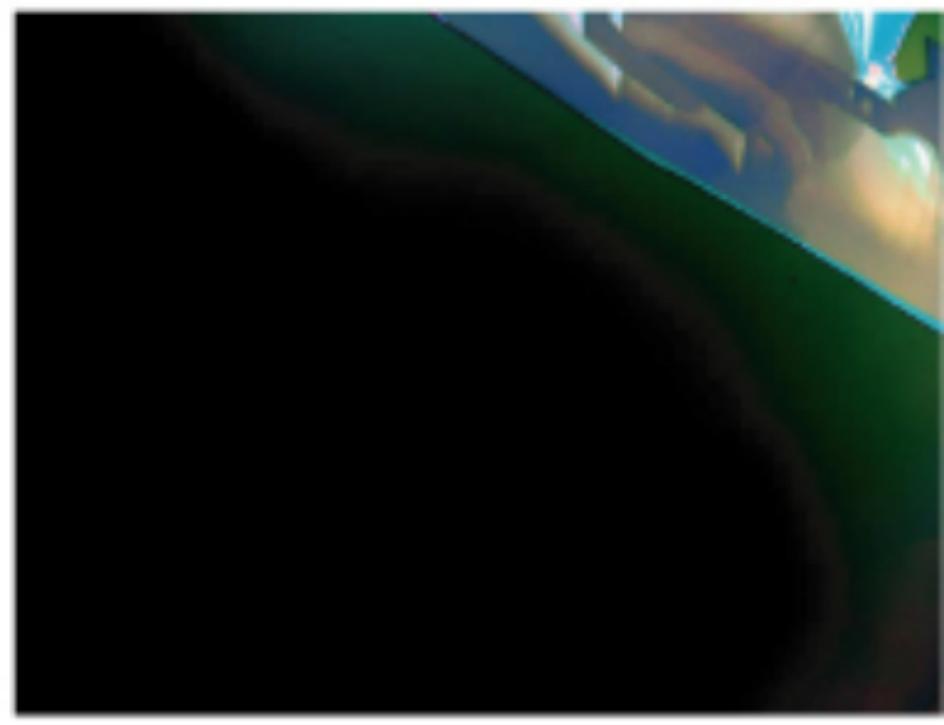

(a)               (b)               (c)

**Supplementary Figure 1**

The diagram shows a schematic of the process used for depositing thin films. Deposition occurs when the solvent evaporates and solution flows into the region of the meniscus in order to maintain the shape of the meniscus. As the liquid level is reduced, the dissolved material condenses into a solid (MS Word Document; 24 kB).

**Supplementary Figure 2**

Schematic representation of the herringbone packing of Anthracene ($C_{14}H_{10}$) looking down along the *c*-axis. The long axis of the molecule is oriented along the *c*-axis, which is directed out of the page in the diagram (MS Word Document; 23 kB).

**Supplementary Figure 3**

The figure shows two optical micrographs of a thin film deposited from a solution containing silylethynyl-substituted derivative of pentacene dissolved in ethyl acetate (MS Word Document; 227 kB).

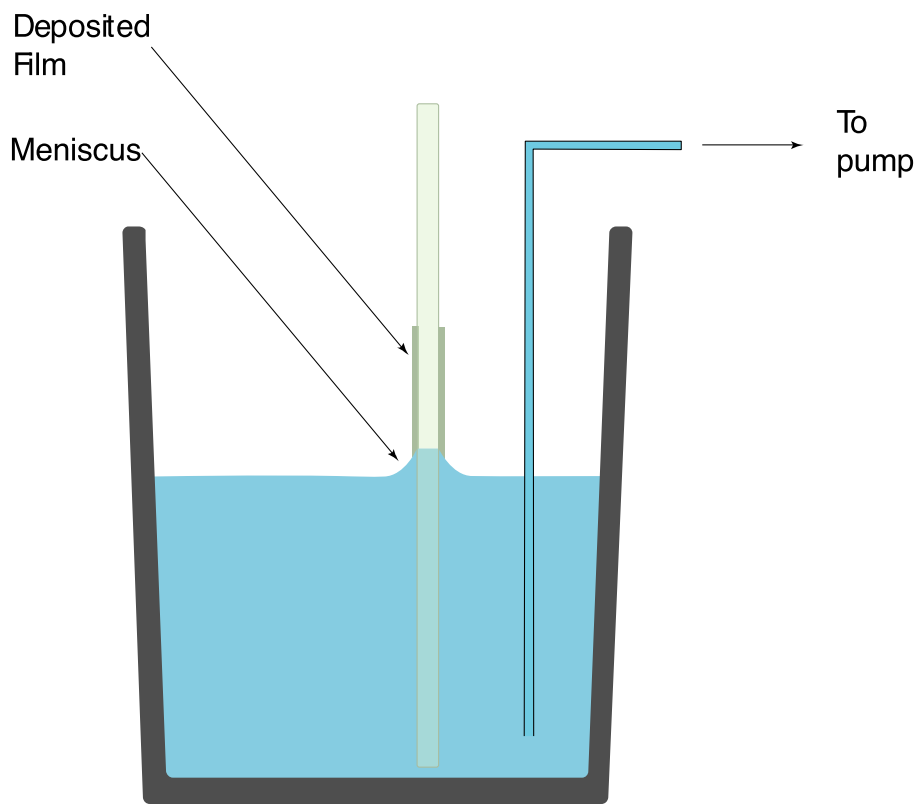

**Deposited Film**

**Meniscus**

To pump

**Supplementary Figure 1** shows an apparatus for depositing thin films using the new method. Components include: (i) a container, which may include an open top; (ii) a solution containing the small molecule to be crystallized; and (iii) a substrate with flat surfaces. The mechanism to pump the solution is not shown. The effect occurs when the solvent evaporates and solution flows into the region of the meniscus in order to maintain the shape of the meniscus. Dissolved molecules are carried along, and concentrate in that region. As the liquid level is reduced, either by evaporation or by external control, the dissolved material condenses into a solid. The deposited film subsequently "seeds" growth at the new position of the meniscus, and the process continues, ultimately coating the entire wetted portion of the substrate with a uniform crystalline layer.

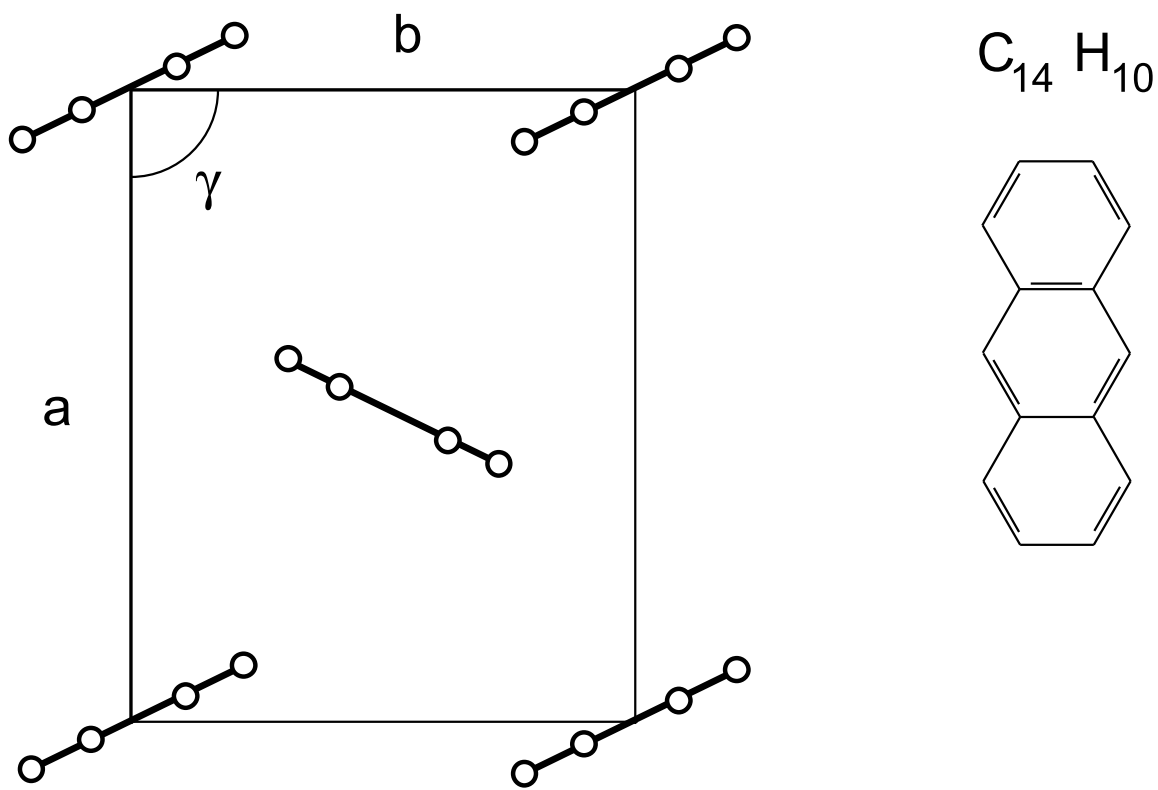

$C_{14} H_{10}$

**Supplementary Figure 2** shows a schematic representation of the herringbone packing of Anthracene ($C_{14}H_{10}$) looking down along the *c*-axis. The structure is composed of layers of molecules stacked along the the c-direction with the "herringbone" packing within each layer. Anthracene has a monoclinic structure with lattice constants *a* = 8.561 Å, *b* = 6.036 Å, *c* = 11.163 Å and β = 124° 42'.

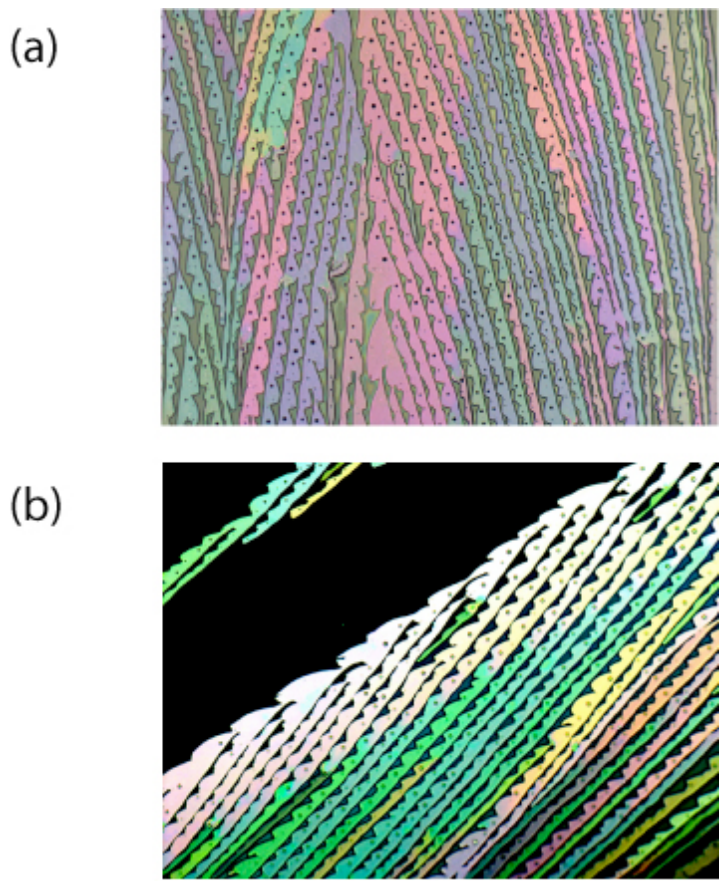

**Supplementary Figure 3** shows two 1.75 mm by 1.25 mm optical micrographs of the same area of a thin film of silylethynyl-substituted pentacene that has been deposited onto a glass slide.  Image (a) shows the film in bright field, while image (b) is with crossed polarizers, and is rotated by 55º with respect to (a).  Areas with different domain orientations in (a) are seen to exhibit different contrast under polarizing conditions, thus confirming that their crystallographic orientations are correspondingly rotated. This film was grown without solvent pumping.  However, a variation of the procedure was employed where the solution was held at 70º C, accelerating the evaporation of the solvent.  The accelerated lateral growth rate led to the formation of rowlike structures rather than continuous domains.  This is in contrast to the continuous domains observed for deposition at room temperature under similar conditions, as exemplified in Fig. 4 of the main text.